\documentclass[aip,pop,amsmath,amssymb,preprint]{revtex4-1}
\usepackage{graphicx}              
\usepackage[caption=false]{subfig} 
\usepackage{dcolumn}               
\usepackage[hidelinks]{hyperref}   
\hypersetup{
  colorlinks = true,               
  urlcolor   = blue,               
  linkcolor  = blue,               
  citecolor  = blue                
}
\usepackage[]{units}               
\usepackage{bm}                    
\newcommand{\sgn}[1]{\operatorname{sgn}\left(#1\right)}
\newcommand{\abs}[1]{\left|#1\right|}
\newcommand{\ifrac}[2]{{#1}/{#2}}
\newcommand{\fpartial}[2][]{\frac{\partial #1}{\partial #2}}
\newcommand{\ipartial}[2]{\ifrac{\partial #1}{\partial #2}}
\newcommand{\spartial}[1]{\partial_#1}
\newcommand{\range}[2]{#1 \textnormal{--} #2}
\newcommand{\dd}[1]{\operatorname{d}\!#1}
\newcommand{\ie}{\emph{i.e.}}
\newcommand{\eg}{\emph{e.g.}}

\date{\today}

\begin{document}
  \title{Three species one-dimensional kinetic model for weakly ionized plasmas}
  \author{J.~Gonzalez}
  \email{jorge.gonzalez@upm.es}
  \author{J.M.~Donoso}
  \author{S.P.~Tierno}
  \affiliation{Department of Applied Physics, Escuela T\'{e}cnica Superior de Ingenier\'{i}a Aeron\'{a}utica y del Espacio, Universidad Polit\'{e}cnica de Madrid, 28040 Madrid, Espa\~{n}a}
  
  \begin{abstract}
    A three species one-dimensional kinetic model is presented for a spatially homogeneous weakly ionized plasma (WIP) subjected to the action of a time varying electric field.
    Planar geometry is assumed, which means that the plasma dynamics evolves in the privileged direction of the field.
    The energy transmitted to the charges is be channelized to the neutrals thanks to collisions and impacting the plasma dynamics.
    Charge-charge interactions have been designed as a one-dimensional collision term equivalent to the Landau operator used for fully ionized plasmas.
    Charge-neutral collisions are modelled by a conservative drift-diffusion operator in the Dougherty's form.
    The resulting set of coupled drift-diffusion equations is solved with the stable and robust Propagator Integral Method (PIM).
    This method feasibility accounts for non--linear effects without appealing to linearisation or simplifications, providing conservative physically meaningful solutions.
    It is found that charge-neutral collisions exert a significant effect since a quite different plasma dynamics arises if compared to the collisionless limit.
    In addition, substantial differences in the system evolution are found for constant and temperature dependent collision frequencies cases.
  \end{abstract}
  
  \maketitle
  
  \section{Introduction}\label{sec:introduction}
    Weakly ionized plasmas are usually described by fluid models.
    However, a better understanding of the basic dynamics involved in momentum and energy transfers among plasma species in presence of self--consistent or imposed electromagnetic fields would require kinetic models for a feasible local description.
    The resulting highly non--linear equations are usually arduous to solve analytically, even numerically, without further simplification or linearisation procedures \cite{bittencourt2013fundamentals,schram1991kinetic,kumar1980kinetic,riemann2003kinetic}.
    
    This solution methods can result in non--physical or inaccurate descriptions which could lead to incomplete models able to explain some complex phenomena, as the plasma-wall interaction.
    For instance, the analysis of the interaction between plasma and emissive or collecting walls is a problem often studied by traditional fluid descriptions \cite{takamura2004space,tierno2016existence} but also dealt from a kinetic point of view \cite{schwager1993effects,sheehan2013kinetic,rizopoulou2014electron,langendorf2015effect} by proposing pre-defined velocity distribution functions not directly obtained by solving specialized kinetic equations.
    Many works do not account for collisions or other effects that may become relevant for the plasma dynamics.
    Rarely a work is found where an approach offers Boltzmann and Vlasov equations to derive the distribution functions of the plasma species, as the one presented in Ref.~\onlinecite{chalise2015self}.
    
    Typical cold plasmas are weakly ionized, \ie, the ratio between electrons or ions and neutrals densities lies in the range of $ \left(\range{10^{-8}}{10^{-4}}\right) $.
    These plasmas cannot be always completely described in the non--collisional regime under the approximation of an immobile high density neutral population.
    It is quite usual to disregard the charge-neutral interaction for a local kinetic analysis due to the small collision cross-section and the null action of electric fields on the neutrals.
    Although the collisionless approximation can be valid for hot plasmas, cold weakly ionized plasma models should require to account for collisions to properly describe the evolution of the system
    Departures from equilibrium are intrinsic in this kind of plasmas, mainly due to the presence of electromagnetic fields responsible for the formation of sharp structures, as filaments, that should be analysed under a theoretical frame involving particles interactions.
    
    Therefore, the collective behaviour of the weakly ionized plasma (WIP) in devices as, for example, a plasma thruster, may strongly depend on the charge-neutral interaction.
    These interactions can be significant because the energy transferred to the charged species by the electric field can be channelled to the neutrals due to collisions processes, modifying the distribution functions with respect to typical Maxwellian.
    This process gives rise to a significant impact on the electrons and ions dynamics which are not precisely described by the sole action of an external electric field \cite{levko2015influence,masoudi2011effects,choquet2007hierarchy}.
    Consequently, for some weakly ionized plasmas holding electric currents some models would lead to a non reliable description of the system if collisional effects are not properly considered, even in situations close to thermodynamic equilibrium.
    
    As a consequence, the macroscopic moments for all plasma species may change, depending on this external force, under time and space scales that also can be different from those derived for systems close to the thermodynamic equilibrium.
    A misleading comprehension of the plasma dynamics may lead, for example, to a mistaken interpretation of the experimental measurements for some plasma diagnosis instruments.
    
    In this work, a three species (neutrals, ions and electrons) 1D kinetic model is proposed to study the dynamics of a weakly ionized plasma when spatially homogeneous densities are assumed and an external electric field drives the system evolution.
    The model consists of three convection-diffusion equations accounting for elastic collisional effects among all plasma species in presence of an external time evolving electric field included in the plasma motion in a self--consistently way.
    To numerically solve these non--linear equations, the Propagator Integral Method (PIM)\cite{gonzalez2014propagator,donoso2006nonlinear,donoso1999short} is used in a 1D open velocity space to compute the time evolving distribution function for each species.
    By means of the properties of this integral method, all non--linear effects introduced without linearisation or simplifications providing a conservative and entropic evolution of the problem.
    Another relevant characteristic of this method is that the same time step can be used to advance the fast (electrons) and slow species (ions and neutrals) without affecting the convergence or stability of the scheme.
    This time step can be relatively large, even a tenth of the characteristic time scale of the system.
    The results may be relevant for further analysis to kinetically describing some typical problems as plasma sheaths in front infinite flat walls\cite{takamura2004space,gyergyek2012saturation}.
    
    This paper is divided as follows.
    In Sec.~\ref{sec:three_species} the kinetic model is presented.
    Sec.~\ref{sec:numerical_results} contains the benchmark problems used to test the model.
    These cases are divided in three subsections: conservation of one species with only self--collisions, influence of collisions in WIP and evolution of the three species system under an electric field.
    Finally, the conclusions of this work are included.
    An appendix with the dimensionless problem and a introductory treatment of the numerical procedure applied is also included.
  
  \section{Three species kinetic model}\label{sec:three_species}
    In plasmas where the ratio between charge and neutrals number densities is small, the dynamics may depend on the charge-neutral interaction.
    This interaction becomes important due to the high density of neutrals, which receives part of the energy transmitted to the charged species by the electric field.
    Moreover, the interplay of several transfer phenomena, governed by quite disparate time scales, demands a kinetic description where collisions have to be properly accounted.

    In this work the use of collision terms having the form of the so-called Fokker--Planck operator for each species of the mixture is proposed.
    In these terms, dynamic friction and diffusion in velocity space are self--consistently included.
    The plasma is assumed to remain spatially homogeneous, which means that the dynamics is ultimately controlled by an external uniform electric field, which strongly dominates the plasma dynamics in one privileged direction.
    This suggests that a one--dimensional velocity space kinetic model suffices to analyse the motion of any distribution along the preferred direction.
    
    Although in weakly ionized plasmas collisions between electrons and neutrals play the fundamental role, all collision terms are included here.
    These contributions are constructed under the binary collision approximation but with peculiar descriptions for both dynamical friction and diffusion processes for each type of interaction.
    Hence, each resulting kinetic equation describes the time evolution of the velocity distribution function $ f_\gamma\left(v,t\right) $, where $ \gamma=0,i,e $ (neutrals, ions and electrons respectively).
    
    Therefore, the kinetic model presented here consists of a system of three non--linear coupled 1D convection-diffusion equations.
    The collision operators are constructed to satisfy the standards of conservative properties accounting for elastic collisions.
    Inelastic collision terms, as ionization or recombination process could be included in a straightforward manner as non--homogeneous source-sink terms \cite{donoso2006nonlinear,gonzalez2014propagator}.
    
    The kinetic equation for a time evolving distribution function for a charged species $ \alpha=e,i $ has the form
    \begin{equation}
      \fpartial[f_\alpha]{t} + \frac{q_\alpha E(t) }{m_\alpha}\fpartial[f_\alpha]{v} = \sum_\gamma \left(\fpartial[f_\alpha]{t}\right)_{\alpha\gamma},\label{eq:fp_c}
    \end{equation}
     where $ m_\alpha $ and $ q_\alpha $ are the mass and charge of the species $ \alpha $ respectively, and $ E(t) $ is the electric field.
    The same form also stands for neutrals by setting $ q_\alpha = 0 $.
    The term $ \left(\ipartial[f_\alpha]{t}\right)_{\alpha\gamma} $ symbolizes the rate of change of the distribution function due to collisional effects between species of kinds $ \alpha $ and $ \gamma $.
    Each of these exchanges are modelled by a drift-diffusion operator, similar to the usual Fokker--Planck collisional one, as 
    \begin{equation}
      \left(\fpartial[f_\alpha]{t}\right)_{\alpha\gamma} \!\!\!\! =-\fpartial{v}\left\lbrace A_{\alpha\gamma} - \fpartial{v}D_{\alpha\gamma} \right\rbrace f_\alpha\label{eq:A_D_coll} .
    \end{equation}
    The parameters $ A_{\alpha\gamma}=A_{\alpha\gamma}\left(f_\alpha, f_\gamma, v, t\right) $ and $ D_{\alpha\gamma} = D_{\alpha\gamma}\left(f_\alpha, f_\gamma, v, t\right) $ are referred to the non--linear convection and diffusion coefficients, respectively.
        
    First, the charge-charge collision term is constructed for the one velocity dimension case by taking as a reference the complete well-known Landau collisional operator\cite{bittencourt2013fundamentals,frank2005nonlinear,pezzi2015collisional} in an spatially homogeneous plasma.
    A physical realisation of this model could be related, for instance, to the description of a plasma between two planar metal walls, where an electric field exists\cite{gyergyek2012saturation,schwager1993effects}.
    In such a system, the planar geometry leads to plasma species distribution functions depending only on the velocity component $v$ lying in the privileged direction established by the electric field.
    Since a test particle of mass $m_\alpha $ and velocity $ v $ is scattered by particles of velocity $ v' $ with distribution function $f_{\beta}\left(v',t\right) $, 
    a small change of the particle velocity $\Delta v $ in the preferred direction may be considered as a result of an average Coulombian interaction force among charges.
    Thence, the dynamical frictional effect experienced by $m_\alpha$ can be phenomenologically modelled as an effective force of uniform intensity opposite to the relative velocity $v-v'$.
    This contribution can be understood as a Coulomb's like law for dry friction, a case also studied in the theoretical frame of Brownian motion\cite{touchette2010brownian}.
    The cumulative effect of many interactions gives rise to the drift coefficient $ A_{\alpha \beta } $ which is related to the change of the expectation value of $ \Delta v$ per unit of time ($ \ifrac{\langle \Delta v \rangle}{\Delta t} $) as 
    \begin{equation}
      A_{\alpha \beta } = \frac{\langle \Delta v \rangle }{\Delta t}= \!\! - \mu_{\alpha\beta} \! \left( 1 \! + \! \frac{m_\alpha}{m_\beta}\right) \!\!\! \int\limits_{-\infty}^{\infty} \!\!\! \sgn{v-v'} f'_\beta \dd{v'}.\label{eq:A_cc}
    \end{equation}
    Here, $\sgn {\cdot} $ is the sign function with $ \sgn{0}=0 $, primes over a distribution function mean $f'_\gamma=f_\gamma\left(v',t\right) $ and $ \mu _{\alpha\beta}$ is a parameter related to plasma properties.
    Besides the friction force, the test charge is subjected to random fluctuating forces of stochastic nature, responsible of the $f_\alpha$ spreading in velocity space.
    This diffusive behaviour is computed by the coefficient $ D_{\alpha \beta } $, related to the average value of $ \ifrac{( \Delta v)^2}{2} $ per unit of time ($ \ifrac{\langle ( \Delta v)^2 \rangle}{2 \Delta t}$).
    Assuming that $\Delta v$ is of order $v-v'$ meanwhile $\Delta v/\Delta t$ is proportional to the friction term $ \sgn{v-v'}$, a diffusion coefficient proportional to the average value of $(v-v') \sgn{v-v'}= \abs{v-v'}$ is proposed
    \begin{equation}
      D_{\alpha \beta } = \frac{\langle (\Delta v)^2 \rangle }{2 \Delta t}= \mu_{\alpha\beta} \int\limits_{-\infty}^{\infty} \abs{v-v'} f'_\beta \dd{v'},\label{eq:D_cc}
    \end{equation}
    assuming that $ \int \abs{v} f_ \beta dv $ remains finite.
    With this selection, the resulting 1D collisional term for charge--charge interaction behaves as the complete plasma physics Fokker--Planck--Landau operator, in fact, it can be easily checked that it provides a well posed conservative collision operator for a one dimensional plasma.
    Moreover, similar differential relations for the drift and diffusion coefficients fulfilled by the Fokker--Planck--Landau operator are also satisfied.
    In particular, making use of the relation $\spartial{v}\sgn{v-v'} = 2 \delta(v-v')$ the properties
    \begin{eqnarray}
      A_{\alpha \beta }             & = & - \left(1+\frac{m_\alpha}{m_\beta}\right)\fpartial{ v} D_{\alpha \beta} \\
      \fpartial{v} A_{\alpha \beta} & = & -2 \mu_{\alpha\beta} \left( 1+\frac{m_\alpha}{m_\beta}\right) f_{\beta},
    \end{eqnarray}
     are obtained.
    These are equivalent to the relations held by the divergences of the diffusion tensor and the drift vector for the Fokker--Planck--Landau operator\cite{donoso2006nonlinear,schram1991kinetic}.
    
    The parameter $\mu_{\alpha\beta}$ does not alter the conservation properties of the operator, its value has been inferred from the equivalent one appearing in the complete Landau operator for fully ionized plasmas.
    Particularly, $\mu_{\alpha\beta}=\ifrac{4\pi\lambda_{\alpha\beta} q_\alpha^2 q_\beta^2}{m_\alpha^2 V_{th_\alpha}^2} $ which is a parameter related to collision frequencies and energy and momentum transfers due to charge interactions.
    Thus, $ \lambda_{\alpha\beta} $ is the Coulomb logarithm, $ V_{th_\beta}=\sqrt{\ifrac{k T_\beta}{m_\beta}} $ is the thermal velocity of the species $ \beta $ and $ k $ is the Boltzmann constant.
    The conservation of norm, momentum and energy of the whole system is satisfied since the complementary collision parameter verifies $ \mu_{\beta\alpha}=$ $\mu_{\alpha\beta} \left(\ifrac{m_\alpha}{m_\beta}\right)^2 $.
    
    Along the theoretical lines established to describe the interaction between charges, the charge-neutral and neutral-neutral contributions 
    to the collisional rates are also constructed in the form of conservative drift-diffusion operators.
    The effects of collisions with neutral particles should be, again, described by both dynamical friction and velocity space diffusion in a self--consistent way.
    Due to the fact that in the mixture, the neutrals distribution does not deviate drastically from a Maxwellian, a multi-species Dougherty collision operator\cite{anderson2007eigenfunctions,dougherty1967model2} in the Fokker--Planck is proposed.
    The drift and diffusion coefficients are implicitly defined in the final expression 
    \begin{eqnarray}
      &&\left(\fpartial[f_\gamma]{t}\right)_{\gamma 0} \!\!\!\! =-\frac{\nu_{\gamma 0}}{n_0}\fpartial{ v}\left\lbrace-\int\limits_{-\infty}^{\ \infty} \left(v-v'\right) f_0' \dd{v'} \right.\nonumber\\*
      &&\left.       -\frac{1}{n_\gamma} \fpartial{v}\int\limits_{-\infty}^{\ \infty}\int\limits_{-\infty}^{\ \infty} \frac{\left(v-v'\right)^2}{2} f_0' f_\gamma \dd{v'} \! \dd{v} \right\rbrace f_\gamma, \label{eq:c-n_coll}
    \end{eqnarray}
     for the rate of change of $f_\gamma$ due to the interaction of species $\gamma$ with neutrals.
    Now, the parameter $ \nu_{\gamma 0}=n_0 \sigma_{\gamma 0} V_{th_\gamma} $ is related to typical frequencies of each collisional process between species $\gamma$ and neutrals\cite{huba2004nrl}.
    Hence, $ \sigma_{\gamma 0} $ is a cross-section measure for charge-neutral and neutral-neutral interaction.
    It is important, however, to mention here that under a microscopic point of view the real collision frequency is velocity dependent, as in the case of charge-charge interaction.
    
    In view of Eq.~(\ref{eq:c-n_coll}), it is clear that the convective contribution stands for an average viscosity force which is, the result of the cumulative effects of a microscopic friction force, proportional to the relative velocity $ v - v' $, between a test particle with velocity $ v $ and a neutral particle of velocity $ v' $.
    The ensemble of neutrals conforms a medium characterized by a velocity distribution $ f_0\left(v',t\right) $, close to a Maxwellian distribution, with density $ n_0 $ much higher than $n_e$ and $n_i$.
    This fact justifies why the charges, and neutrals as well, experience a viscous friction force stated by Eq.~(\ref{eq:c-n_coll}).
    On the other hand, the diffusion coefficients only depend on macroscopic magnitudes due to the double integral over primed variables stated above.
    The drift and diffusion coefficients included in Eq.~(\ref{eq:c-n_coll}) ensure energy, momentum and norm conservation thanks to the reciprocity relations $ \nu_{0 \gamma}= \nu_{\gamma 0}\ifrac{m_\gamma}{m_0} $.
    Thence, this operator is also fully conservative and it leads to a Maxwellian distribution function when it acts as a self--collision term for a one species system in absence of external forces and inelastic terms.
    The characteristic evolution frequencies of this Dougherty operator do no differ so much from the corresponding ones for a complete Fokker--Planck--Landau operator, as it was analysed in Ref.~\onlinecite{pezzi2015collisional}.
    
    As for Eq.~(\ref{eq:fp_c}) for $ q_\alpha=0 $, the velocity distribution function for the neutrals obey a convection-diffusion equation of the form
    \begin{equation}
      \fpartial[f_0]{t}=\sum\limits_{\gamma}\left(\fpartial[f_0]{t}\right)_{0 \gamma},\label{eq:fp_0}
    \end{equation}
     which accounts for the energy transferred from the charged species to neutrals.
    The collision terms follow Eq.~(\ref{eq:c-n_coll}) with the indexes $ \gamma $ and $ 0 $ interchanged.
    It can be seen that although the neutrals do not directly experience the action of electromagnetic forces, the velocity distribution $ f_0 $ may evolve due to collisional effects with charged species.
    As a consequence, the neutrals acquire a small amount of the electrical energy transferred to the charges\cite{donoso2015numerical}.
    This means that all the species feel directly or indirectly the perturbation induced by an electric field.
    
    It is important to mention that the collision parameters $ \mu_{\alpha \beta} $ and $ \nu_{\gamma 0} $ depend on the temperature through the thermal velocity, the cross-sections and the Coulomb logarithm, introducing a new source of non--linearities.
    In the numerical solution, these parameters must be computed at each time step to take into account the different exchanges among the particles when their temperatures vary.
    The procedure to solve convection-diffusion equations using the PIM method can be found in previous works \cite{donoso2015numerical,gonzalez2014propagator,donoso2006nonlinear,donoso1999short,wehner1987numerical3} and it is not largely presented here.
    However, a introductory analysis of the numerical algorithm, as well as the dimensionless problem, are presented in Appendix ~\ref{appx:numerical_simulation} for a self--consistent reading of this work.
    
  \section{Numerical results}\label{sec:numerical_results}
    In this section, three benchmark problems are solved using the PIM to test the kinetic model previously introduced.
    First, only self--collisions are accounted to study how the terms described in the previous section behave individually, \ie, when they act on a distribution function.
    Then, to analyse the influence of collisions, a comparison between the evolution of a non--collisional plasma and the three species model is shown.
    Next, the complete kinetic model presented in Sec.~\ref{sec:three_species} is computed with a time variable electric field.
    In addition, a comparison between the previous case with constant and temperature dependent collision parameters $ \mu_{\alpha \beta} $ and $ \nu_{\gamma 0} $ is performed.
    
    \subsection{One species conservative collision term}\label{subsec:one_species}
      The collision terms  described in this paper are analytically conservative in norm, momentum and energy.
      To properly include these terms in more complex simulations, the numerical solution of simple dimensionless problems accounting only for self--collisions should also conserve these macroscopic variables.
      
      For the mutual charge-charge interaction, \ie, if $ \alpha=\beta $, the collision term reads
      \begin{eqnarray}
        \left(\fpartial[f]{t}\right)_{\alpha\alpha} \!\!\!\! & = & - 
        \fpartial{v}
          \left\lbrace -2             \int\limits_{-\infty}^{\ \infty} \sgn{v-v'} f' \dd{v'} \right.         \nonumber\\*
        &&\left.       -\fpartial{ v} \int\limits_{-\infty}^{\ \infty} \abs{v-v'} f' \dd{v'} \right\rbrace f \label{eq:c-c_coll_1S},
      \end{eqnarray}
       where the parameter $ \mu_{\alpha \alpha} $ has been set to $ 1 $.
      The initial condition is a Maxwellian distribution function with dimensionless density, mass and temperature equal to unity.
      The main results of this test problem are shown in Fig.~\ref{fig:CC_1S}.
      \begin{figure*}
        \subfloat[\label{fig:CC_1S_a}]{\includegraphics{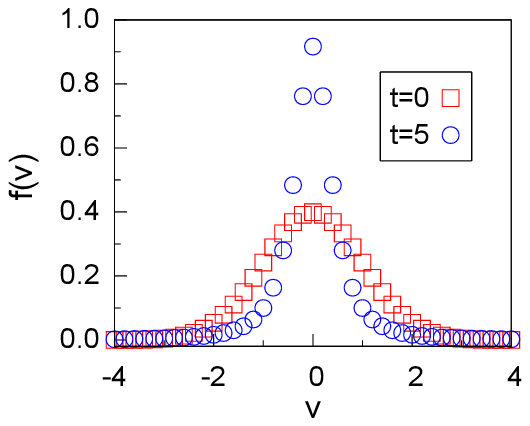}}
        \subfloat[\label{fig:CC_1S_b}]{\includegraphics{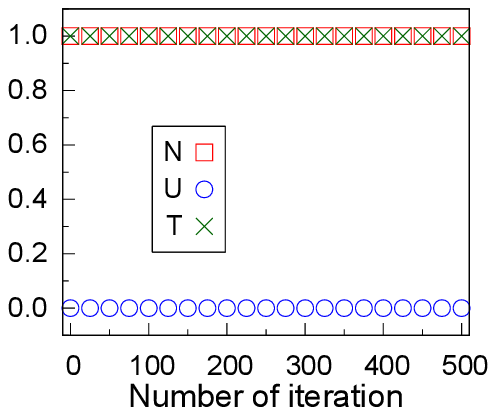}}
        \subfloat[\label{fig:CC_1S_c}]{\includegraphics{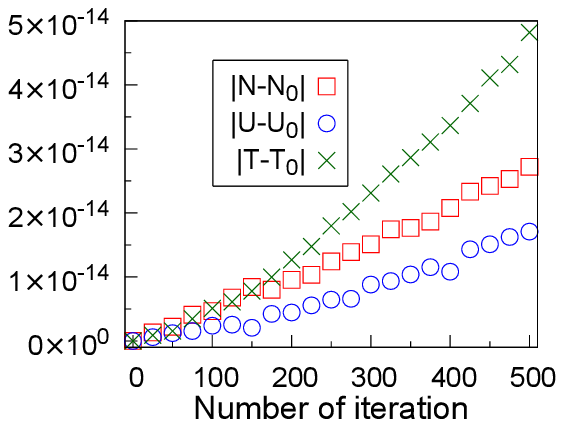}}
        \caption{\label{fig:CC_1S}Solution for the charge-charge self--interaction.
                                  The time step is $ \tau=0.01 $.
                                  (a)~Distribution function for initial condition (red squares) and at the end of the simulation (blue circles).
                                  (b)~Evolution in time of the macroscopic moments.
                                  (c)~Error respect to the initial condition.}
      \end{figure*}
      When this collision term acts alone, the solution evolves to a distribution function with power-law tails\cite{gonzalez2014propagator} (Fig.~\ref{fig:CC_1S_a}), but preserving the initial temperature, density and velocity (Fig.~\ref{fig:CC_1S_b}).
      The moments of the distribution function are computed through the relations
      \begin{eqnarray}
        \left(\begin{array}{c} N\\ U\\ T\\ \end{array}\right)=\int\limits_{-\infty}^{\ \infty}\left(\begin{array}{c} 1\\ v\\ \tfrac{1}{2}\left(v - u\right)^2\\ \end{array}\right) f\left(v,t\right) \dd{v}
      \end{eqnarray}
       where $ u=\ifrac{U}{N} $.
      If the difference of these macroscopic quantities respect to the initial condition are analysed (Fig.~\ref{fig:CC_1S_c}), a very small difference in the numerical results appears.
      The order of these errors ($ \sim 10^{-14} $) is small enough to consider the numerical solution unperturbed, even for a large number of iterations.
      
      The same procedure is applied for the collision term of Eq.~(\ref{eq:c-n_coll}).
      If only collisions between the neutral particles are accounted ($ \gamma=0 $), the mutual collision term becomes
      \begin{eqnarray}
         &&\left(\fpartial[f]{t}\right)_{0 0} \!\!\!\! = -
           \fpartial{ v}
           \left\lbrace  -\int\limits_{-\infty}^{\ \infty}  \left(v-v'\right) f' \dd{v'} \right.\nonumber\\*
         &&\left.        -\fpartial{v}\int\limits_{-\infty}^{\ \infty}\int\limits_{-\infty}^{\ \infty} \frac{\left(v''-v'\right)^2}{2} f' f'' \dd{v'} \dd{v''} \right\rbrace f\label{eq:c-n_coll_1S},
      \end{eqnarray}
       where, again, the collision frequency $ \nu_{0 0} $, density, mass and temperature equal $ 1 $.
      Figures~\ref{fig:CN_1S_a} and~\ref{fig:CN_1S_b}
      \begin{figure*}
        \subfloat[\label{fig:CN_1S_a}]{\includegraphics{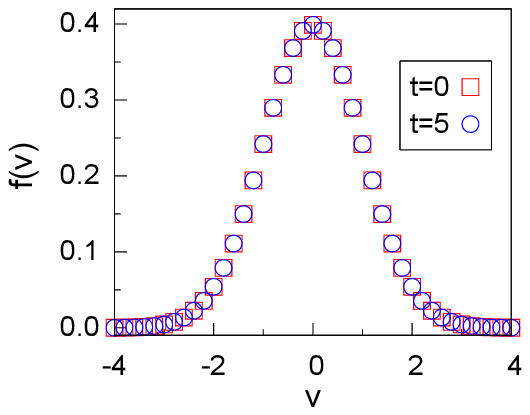}}
        \subfloat[\label{fig:CN_1S_b}]{\includegraphics{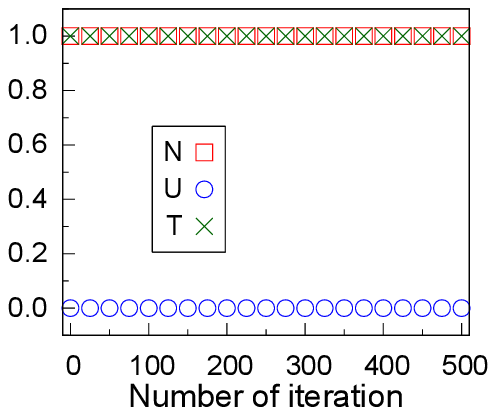}}
        \subfloat[\label{fig:CN_1S_c}]{\includegraphics{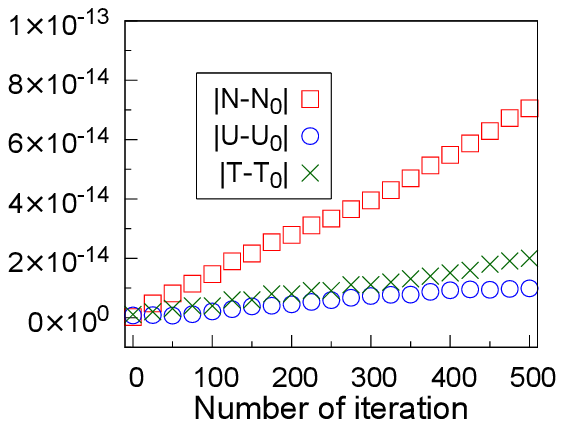}}
        \caption{\label{fig:CN_1S}Solution for the neutral-neutral interaction problem.
                                  The time step is $ \tau=0.01 $.
                                  (a)~Distribution function for initial condition (red squares) and at the end of the simulation (blue circles).
                                  (b)~Evolution in time of the macroscopic moments.
                                  (c)~Error of the moments respect to the initial condition.}
      \end{figure*}
       show how the initial Maxwellian distribution remains almost constant, since small numerical errors (presented in Fig.~\ref{fig:CN_1S_c}) appear.
      
      These test evidence that the PIM is conservative and only numerical errors related to the machine's precision and the mesh finiteness appear in the simulations.
      
    \subsection{Influence of collisions}\label{subsec:influence_neutrals}
      To test the effect and relevance of the energy and momentum transfers due to collisions in weakly ionized plasmas, a comparison between a non--collisional and the three species kinetic model is presented here.
      The system of equations that describes the evolution of a non--collisional plasma derives from Eq.~(\ref{eq:fp_c}) and~(\ref{eq:fp_0}) without collision terms.
      Thus, the Vlasov equations under space homogeneity are,
      \begin{equation}
        \fpartial[f_e]{t} - \frac{\abs{q_e} E}{m_e} \fpartial[f_e]{v}=0,
        \fpartial[f_i]{t} + \frac{\abs{q_i} E}{m_i} \fpartial[f_i]{v}=0.\label{eq:vlasov}
      \end{equation}
      The fluid velocities for the charged species can be analytically calculated, if the electric field is given as a time varying function, as
      \begin{equation}
        u_\alpha\left(t\right) = \mp \frac{\abs{q_\alpha}}{m_\alpha}\int\limits_{t_0}^{t} E\left(t'\right) \dd{t'} + u_\alpha\left(t_0\right)\label{eq:u_NCP},
      \end{equation}
       where $ t_0 $ is the initial instant, sign ($ - $) corresponds to electrons and ($ + $) to ions.
      The species norm and temperature do not change for this non--collisional homogeneous plasma.
      Neither the macroscopic momentum nor the distribution function of neutrals change ($ \ipartial[f_0]{t}=0 $) because no interaction with the electric field or with any other species appear.
      
      To check the influence of the collisions, a simple test-case is performed.
      In this test, the electric field has the form
      \begin{equation}
        E\left(t\right)=
          \begin{cases}
             \ \ E_0 & \textnormal{if } t   <  \unit[2.5]{ms} \\
             -   E_0 & \textnormal{if } t \geq \unit[2.5]{ms}
           \end{cases}\label{eq:E_Abrupt}.
      \end{equation}
       where $ E_0=\unitfrac[0.1]{V}{m} $, which is in the order of magnitude characteristic in a WIP.
      
      In order to emulate an Argon weakly ionized plasma, the following parameters are employed with Maxwellian distribution functions as initial conditions
      \begin{eqnarray}
         & m_0=m_e+m_i;\,            T_0=0.03 \times T_e;\,  n_0=\unit[10^{13}]{cm^{-3}}\nonumber\\*
         & m_i=72819.6\times m_e;\,  T_i=0.05 \times T_e;\,  n_i=r_i \times n_0         \nonumber\\*
         &                           T_e=\unit[1]{eV};\,     n_e=r_i \times n_0\label{eq:Argon_param},
      \end{eqnarray}
       where $ m_\gamma $, $ T_\gamma $ and $ n_\gamma $ are the mass, temperature and density of the species $ \gamma $, respectively; and the ionization ratio ($ r_i $) is $ 10^{-6} $.
      In addition, the fluid velocities are zero ($ u_\gamma=0 $) and single charge ions are assumed ($ Z = 1 $).
      For the cases presented in this work, the cross-sections for the charge-neutral and neutral-neutral interactions are considered as constant and equal to $ \sigma_{00}=\sigma_{i0}=\unit[10^{-14}]{cm^{-2}};\; \sigma_{e0}=\unit[10^{-16}]{cm^{-2}} $ although these parameters should change with the species energy\cite{mitchner1973partially}.
      The remaining parameters required to compute $ \mu_{\alpha\beta} $ and $ \nu_{\gamma 0} $ are extracted from the literature\cite{huba2004nrl,mitchner1973partially}.
      To properly represent the evolution of the system, collision parameters $ \mu_{\alpha\beta} $ and $ \nu_{\gamma 0} $ are updated at each time step in the simulation.
      
      It can be seen in Eq.~(\ref{eq:Argon_param}) that very disparate time scales appear naturally in the problem due to the huge ratios of temperature, density and mass among the species.
      Classical numerical methods may become unstable leading to non--physical solutions when these ratios or non--linearities appear in the numerical problem.
      The PIM can deal with these disparate scales without introducing instabilities in the numerical solution and keeping a physically meaningful time-evolving solution.
      
      In Fig.~\ref{fig:Inf_0},
      \begin{figure}
        \subfloat[\label{fig:Inf_0_a}Non--collisional plasma]{\includegraphics{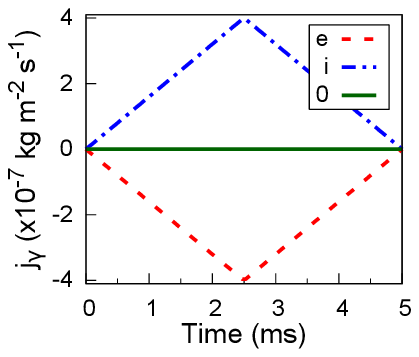}}
        \subfloat[\label{fig:Inf_0_b}Collisional plasma]{\includegraphics{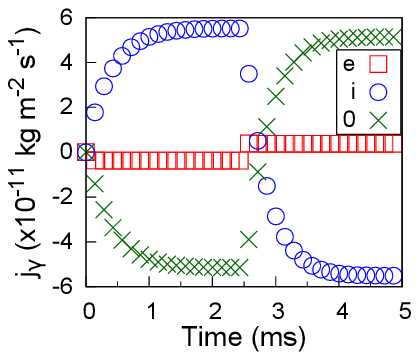}}
        \caption{\label{fig:Inf_0}Evolution of the mass fluxes $ j_\gamma $ for the three species in the cases (a) non--collisional plasma and (b) the three species 1D kinetic collisional plasma.}
      \end{figure}
       the mass fluxes ($ j_\gamma=m_\gamma n_\gamma u_\gamma $) for the non--collisional plasma (Fig.~\ref{fig:Inf_0_a}) and the model described in this work (Fig.~\ref{fig:Inf_0_b}) are presented.
      A complete different dynamics appears when collisions are taken into account.
      In one hand, dynamics without collisions follows Eq.~(\ref{eq:u_NCP}) for the charged species and no change for the neutrals.
      On the other hand, when collisions are accounted, electron mass flux reaches a fast maximum and then remains constant until the electric field changes ($ t=\unit[2.5]{ms} $).
      Ions and neutrals also reach a steady velocity, but a longer time is required due to their high mass.
      
      One surprising result is that neutrals are accelerated even when this species does not directly feel any effect of the electric field.
      This evolution is a result of the energy exchanged to the neutrals by the charged species, which implies that the charge-neutral interaction is important to study the dynamics of WIP.

      Even when the total energy introduced to the system is the same in both, collisional and non--collisional, cases, this energy is redistributed in a different way.
      Mass fluxes reached for charged species are orders of magnitude below the obtained for the non--collisional case.
      This occurs because when collisions are introduced, part of the energy transferred by the electric field is finally channelled to the neutrals, which act as a background that slowdown the charged species.
      
      The different time scales introduced by the disparate ratios are irrelevant to the non--collisional model, where the mass fluxes for electrons and ions evolve symmetrically.
      If collisions are accounted, each species evolve with a different dynamics.
      This means that, in weakly ionized plasmas, collisions should not be neglected for large time scales.
      Also, the distribution function evolution of the neutrals is conditioned by the exchange of energy and momentum with the charged species.
    
    \subsection{Three species evolution under an abrupt electric field}\label{subsec:three_species_abrupt_E}
      Here, the three species kinetic model is tested to the electric field of Eq.~(\ref{eq:E_Abrupt}) for different values of $ E_0 $.
      Moreover, a comparison between the evolution of the system with constant and temperature dependent collision parameters is performed.
      
      In Fig.~\ref{fig:j_Variable_nu},
      \begin{figure*}
        \subfloat[\label{fig:j_Variable_nu_a}Electrons]{\includegraphics{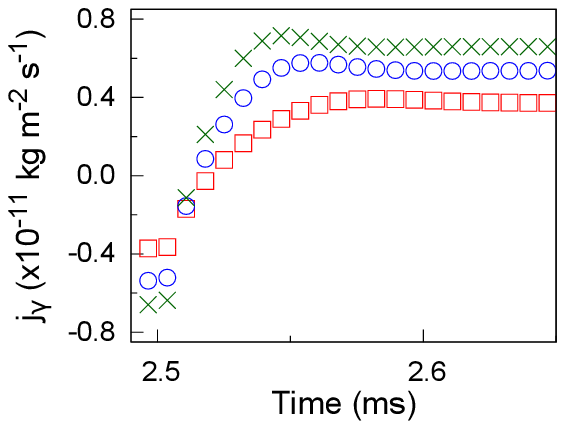}}
        \subfloat[\label{fig:j_Variable_nu_b}Ions]     {\includegraphics{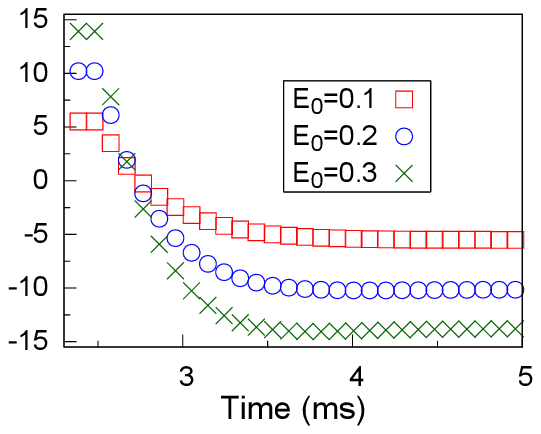}}
        \subfloat[\label{fig:j_Variable_nu_c}Neutrals] {\includegraphics{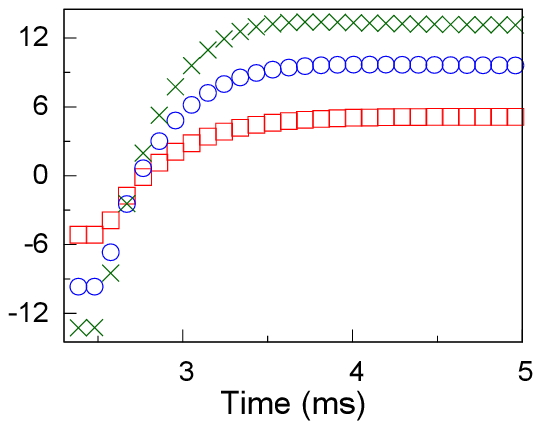}}
        \caption{\label{fig:j_Variable_nu}Mass fluxes for (a) electrons, (b) ions and (c) neutrals.
                                          Electrons have a fast relaxation time due to the high thermal velocity.
                                          Meanwhile, heavy and cold species require more time to recover a constant velocity.}
      \end{figure*}
       a zoomed view of the mass fluxes when the electric field changes ($ t=\unit[2.5]{ms} $) for $ E_0=\unitfrac[0.1\textnormal{, }0.2\textnormal{ and }0.3]{V}{m} $ is shown.
      It is important to remark that this abrupt change in the electric field does not produces oscillations or instabilities in the time evolving numerical solution.
      In these results, different dynamics between the fast (electrons) and the slow (ions and neutrals) species can be apprehended.
      First of all, electrons only require fractions of millisecond to recover a constant velocity.
      On the other hand, ions and neutrals require several milliseconds to reach an steady state.
      These different behaviours are directly related to the exchange rate of energy with neutrals.
      
      As it was indicated in Eq.~(\ref{eq:Argon_param}), electrons have small mass at high temperature, which means they have higher thermal velocity than ions ($ V_{th_e} \gg V_{th_i} $).
      The charge-neutral collision frequency $ \nu_{\gamma 0} $ is directly proportional to the thermal velocity, so that, the rate of electron-neutral exchange is higher than the ion-neutral one.
      This results in a faster relaxation time, even when $ \sigma_{e0} < \sigma_{i0} $.
      Here, relaxation time refers to the time required by one species to recover a constant velocity after the electric field changes its magnitude.
      
      Also, when higher values of $ E_0 $ are applied, differences in the overshoots and relaxation times appear, specially for electrons.
      The overshoot in the mass flux of electrons (Fig.~\ref{fig:j_Variable_nu_a}) grows with the electric field, but electrons recover quickly the constant velocity.
      This means that a higher temperature for the charged species is reached, which results in a higher charge-neutral collision frequency.
      It can be seen in Fig.~\ref{fig:T_Variable_nu},
      \begin{figure}
        \subfloat[\label{fig:T_Variable_nu_a}Electrons]{\includegraphics{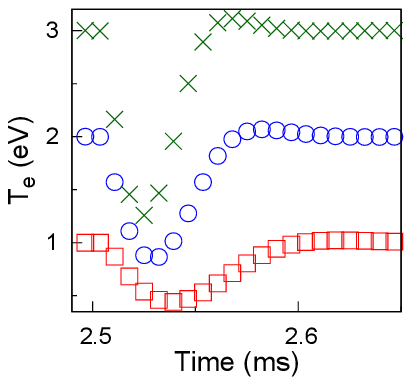}}
        \subfloat[\label{fig:T_Variable_nu_b}Ions]     {\includegraphics{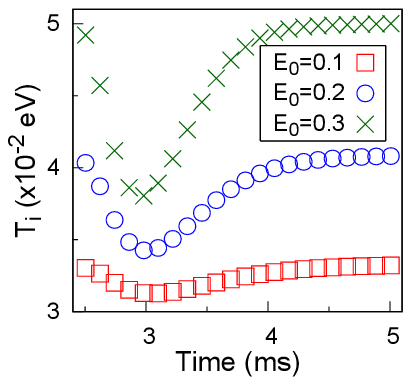}}
        \caption{\label{fig:T_Variable_nu}Temperatures for (a) electrons and (b) ions.
                                          The species recover faster a steady state when high electric field is applied.}
      \end{figure}
       how temperature increases with the electric field.
      As it was explained above, the temperature modifies the collision frequencies through the thermal velocity, resulting in more encounters with neutrals.
      Then, the value of electric field influences not only the velocity of the charges but also their temperature, which ends up modifying the collision parameters and the dynamics of the whole system.
      On the contrary, when the electric field changes, temperature decreases until the electrons reach again a constant velocity, which produces less electron-neutral exchange.
      This allows electrons to reach a fast velocity, until temperature increases again and the excess of energy is finally transferred to neutrals.
      This behaviour can only be obtained if the plasma is modeled in a self--consistently way.
      
      To test the influence of the non--constant collision parameters, the same test cases are performed keeping the initial values of these parameters through all the simulations.
      The mass flux of electrons is studied in Fig.~\ref{fig:comp}
      \begin{figure*}
        \subfloat[\label{fig:comp_a}$ E_0=0.1\,{\unitfrac{V}{m}} $]{\includegraphics{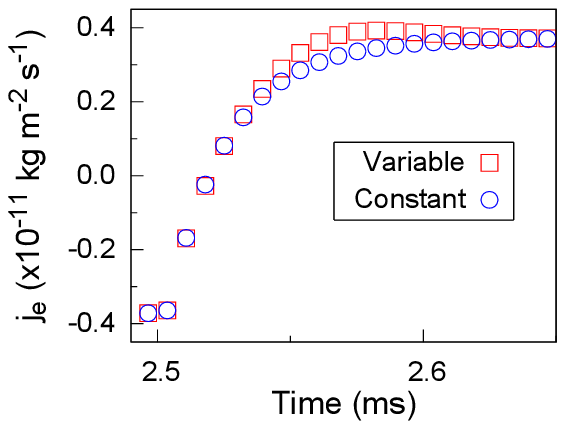}}
        \subfloat[\label{fig:comp_b}$ E_0=0.2\,{\unitfrac{V}{m}} $]{\includegraphics{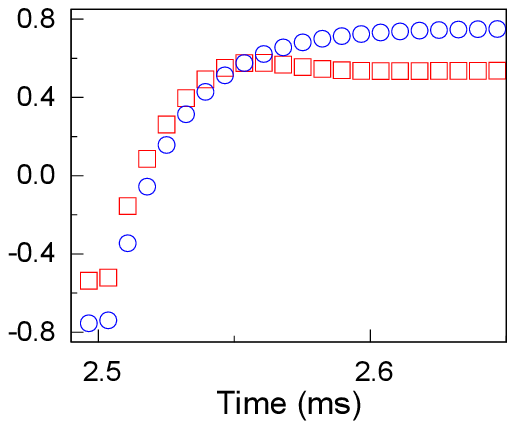}}
        \subfloat[\label{fig:comp_c}$ E_0=0.3\,{\unitfrac{V}{m}} $]{\includegraphics{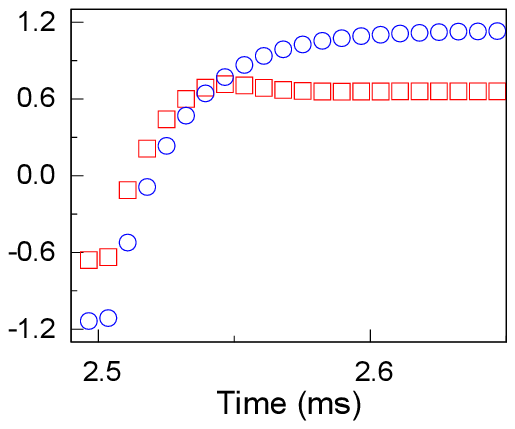}}
        \caption{\label{fig:comp}Comparison of the electron mass flux with variable and constant collision parameters.
                                 Very different dynamics appear for both problems and no overshoot raise with constant collision parameters.
                                 Differences in the evolution of the slow species are also found.}
      \end{figure*}
       for different values of $ E_0 $ for both cases.
      A difference in the relaxation times can be appreciated for constant collision frequencies.
      When collision parameters depend on the species temperature, an overshoot in the electrons mass flux appear due to the decrement in temperature.
      With constant collision frequencies, no change or overshoot in the dynamics of the species appear because, even at low temperature, electrons interact with neutrals at a higher rate that the one that correspond to their temperature.
      
      These results indicate that keeping collision parameters as their initial (or any other constant) value could produce inaccurate results when species temperatures change substantially during the simulations, which can occur, for example, when abrupt changes in the electric field appear.

      To check if the solutions presented in this problem have a physically meaningful evolution in time, the system entropy is now studied.
      The Boltzmann entropy is computed as
      \begin{equation}
        S\left(t\right) = - k \sum_{\gamma} \int f_\gamma\left(v,t\right) \log{f_\gamma\left(v,t\right)} \dd{v}.
      \end{equation}
      The time derivative of $ S\left(t\right) $ is obtained by a simple first order central numerical derivative scheme for the inner time frames, and first order upwind forward and backward for the first and last frames respectively.
      In Fig.~\ref{fig:dS},
      \begin{figure}
        \subfloat[\label{fig:dS_a}Temperature dependent collision parameters]{\includegraphics{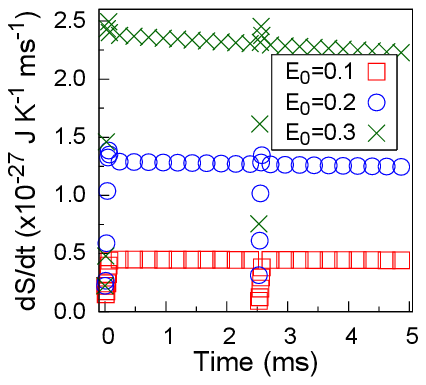}}
        \subfloat[\label{fig:dS_b}Constant collision parameters]{\includegraphics{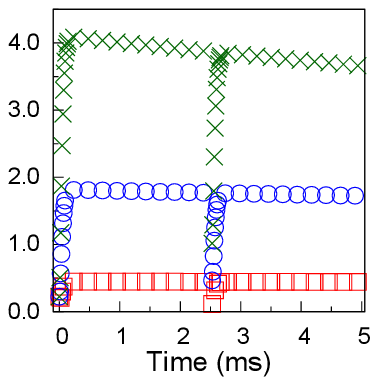}}
        \caption{\label{fig:dS}System entropy derivative for different electric field for (a) variable and (b) constant collision parameters.}
      \end{figure}
       this derivative is presented.
      As it can be seen, this value is always positive, even when an abrupt change in the electric field occurs.
      This means that the numerical solution agrees with the expected thermodynamic evolution of increasing entropy.
    
  \section{Conclusions}\label{sec:conclusions}
    In this work, a one-dimensional kinetic model has been presented to describe the evolution of three species spatially homogeneous weakly ionized plasma.
    A planar geometry has been assumed to design all the elastic collision terms.
    The proposed charge-charge collision term corresponds to a one-dimensional equivalent Fokker-Planck-Landau operator.
    A non--linear Dougherty operator has been adopted for the collisional terms involving neutrals.
    
    The Propagator Integral Method is employed to solve the coupled equations numerically, using the same time step for fast and slow species dynamics.
    This advancing scheme provides a conservative evolution of the system regardless non--linearities involved in the drift-diffusion parameters.
    It is found that if usual linearisation or simplification procedures are applied, \eg, constant collision frequencies, some dynamics could remain hidden, such as the overshoots that appear for fast changes in the electric field.
    Accordingly to the properties of the collision operators, time evolving velocity distributions functions are obtained, independently of the species mass, density and temperature ratios.
    An entropic consistent evolution of the system is also obtained for all values of the applied electric field, even do this external force changes abruptly with respect to time.
    Furthermore, these drastic changes in the electric field do not affect the consistency of the numerical method, which produces a smooth transition in the time depending macroscopic variables.
     
    We can conclude that it is important to solve the WIP kinetic equations in a self--consistent way to provide a physical meaningful time evolving solution.
    In particular, we stress that collisional effects, specially charge-neutral interactions, can exert a significant influence in the dynamics of weakly ionized plasmas.
    Thus, the high density of neutrals produces a viscous-like effect on the charged species.
    This effect transfers part of the energy transmitted by the electric field to the charges to the neutral population.
    We remark that such an effect is impossible to be recovered with a non--collisional plasma kinetic approximation.
    Additionally, the effect of the temperature on the collision parameters influences the dynamics of the system.
    This description can be important for a better understanding of results obtained by diagnosis procedures based upon the use of fast sweep signals generating time and space depending electric fields.
    
  \appendix
  \section{Dimensionless numerical simulation}\label{appx:numerical_simulation}
    For the dimensionless problems presented in Sec.~\ref{subsec:one_species}, $ 10000 $ points are used with a grid that goes from $ -250 $ to $ 250 $ dimensionless velocities in the charge-charge case and from $ -25 $ to $ 25 $ in the neutral-neutral problem.
    The long grid for the charge-charge problem is required due to the potential behaviour of the tails, which require more space to decline up to a value that does not perturb the problem.
    The time step for both problems is $ \tau=0.01 $.
    
    To numerically solve the three species kinetic model presented in Sec.~\ref{sec:three_species}, takes into account the following characteristic parameters:
    \begin{eqnarray}
        & n_c     =n_0;\,  m_c=m_e;\, q_c=q_e;\, T_c=T_e; \nonumber \\
        & \sigma_c=\sigma_{e 0}=\unit[10^{-16}]{cm^{-2}};\, v_c=V_{th_e}; \nonumber\\
        & \mu_c   =\sigma_c v_c;\, \nu_c=n_c \mu_c;\, t_c=\nu_c^{-1};\, E_c=\unitfrac[1]{V}{m}.
    \end{eqnarray}
    The choice of the electron-neutral collision as a characteristic parameter provides a good compromise between the high density of neutral particles and high temperature of electrons.
    The collision parameters are now rescaled with respect to the characteristic values $ \tilde{\mu}_{\alpha\beta}=\ifrac{\mu_{\alpha\beta}}{\mu_c} $ and $ \tilde{\nu}_{\gamma 0}=\ifrac{\nu_{\gamma 0}}{\nu_c} $ where $ \tilde{\;} $ represent dimensionless variables.
    The remaining microscopic and macroscopic variables are also rescaled in the same way.
    
    To advance the distribution functions in time, a simple integral scheme\cite{donoso2015numerical, donoso1999short, wehner1987numerical3} in an absence of source-sink terms is used.
    For the problems presented in this work this scheme essentially reads as
    \begin{equation}
      f_\gamma\left(v,t+\tau\right)=\int\limits_{-\infty}^{\ \infty}P_{\tau_\gamma}\left(v,v'\right)f_\gamma\left(v',t\right) \dd{v'},
    \end{equation}
     where
    \begin{equation}
      P_{\tau_\gamma}\left(v,v'\right)=\frac{1}{\sqrt{4\pi D'_\gamma \tau}}\exp{\left(-\frac{\left(v-v'-\tau A'_\gamma\right)^2}{4 D'_\gamma \tau}\right)}
    \end{equation} 
    is the Gaussian distribution, known as a propagator, where $ A_\gamma $ and $ D_\gamma $ are the convection and diffusion parameters that contain all the collision terms coefficients and external forces for the species $ \gamma $, prime over the convection-diffusion parameters means they are evaluated at the source variables and $ \tau $ is the time step.
    This analytical scheme is transformed into a numerical integration one using a simple rectangle method:
    \begin{equation}
      f^{n+1}_{\gamma_j}=\sum_{j=0}^{v_{max}}P^n_{\gamma_{j j'}}f^{n}_{\gamma_{j'}} \Delta v_\gamma,
    \end{equation}
     where $ j $ and $ j' $ are the array indexes, $ v_{max} $ is the maximum number of points, $ n $ is the current iteration, and $ \Delta v_\gamma $ is the grid step for the species $ \gamma $.
    The numerical integrals are performed from $ -\ifrac{L_{v_\gamma}}{2} $ to $ \ifrac{L_{v_\gamma}}{2} $, where $ L_{v_\gamma} $ is the mesh length in the velocity space for the species $ \gamma $.
    To obtain a better representation of the distribution function, different grid lengths are used for each species: $ L_e=100 $, $ L_i=0.05 $, and $ L_0=0.03 $ with $ 5000 $ points for each species.
    The three distribution functions are advanced with the same time step $ \tau=0.01 $.
  
  \begin{acknowledgments}
    This work was funded by the Ministerio de Econom\'ia Ciencia e Innovaci\'on of Spain under Grant ESP2013-41078-R.
    The authors also acknowledge the partial support from Aernnova Engineering S.A.
    The authors acknowledge the computer resources and technical assistance provided by the Centro de Supercomputaci\'on y Visualizaci\'on de Madrid (CeSViMa).
    S.P.~Tierno acknowledges her FPU Grant from the Spanish Ministry of Education (MECD).
  \end{acknowledgments}
  
  \bibliography{shorttitles,bibliography}
\end{document}